\documentstyle[12pt]{article}

\def\dalemb#1#2{{\vbox{\hrule height .#2pt
        \hbox{\vrule width.#2pt height#1pt \kern#1pt
                \vrule width.#2pt}
        \hrule height.#2pt}}}

   \let\d=\delta \let\e=\epsilon
    
 \let\m=\mu    
   \let\f=\phi  
\let\vp=\varphi \let\vep=\varepsilon
       \let\D=\Delta

\let\la=\label  
  
\def\nn{\nonumber} \def\bd{\begin{document}} \def\ed{\end{document}}
\def\ds{\documentstyle} \let\fr=\frac \let\bl=\bigl \let\br=\bigr
\let\Br=\Bigr \let\Bl=\Bigl
\let\bm=\bibitem
\let\na=\nabla
\def\tU{{\widetilde U}}
\let\pa=\partial \let\ov=\overline
\def\ie{{\it i.e.\ }}
\newcommand{\be}{\begin{equation}}
\newcommand{\ee}{\end{equation}}
\def\ba{\begin{array}}
\def\ea{\end{array}}
\def\ft#1#2{{\textstyle{{\scriptstyle #1}\over {\scriptstyle #2}}}}
\def\fft#1#2{{#1 \over #2}}
\def\F#1#2{{ F_{#1}^{(#2)} }}
\def\cF#1#2{{ {\cal F}_{#1}^{(#2)} }}
\def\del{\partial}
\def\R{{\bf R}}
\def\sst#1{{\scriptscriptstyle #1}}
\def\oneone{\rlap 1\mkern4mu{\rm l}}
\def\e7{E_{7(+7)}}
\def\td{\tilde}
\def\wtd{\widetilde}
\def\im{{\rm i}}
\def\bog{Bogomol'nyi\ }
\newcommand{\ho}[1]{$\, ^{#1}$}
\newcommand{\hoch}[1]{$\, ^{#1}$}
\newcommand{\bea}{\begin{eqnarray}}
\newcommand{\eea}{\end{eqnarray}}
\newcommand{\ra}{\rightarrow}
\newcommand{\lra}{\longrightarrow}
\newcommand{\Lra}{\Leftrightarrow}
\newcommand{\ap}{\alpha^\prime}
\newcommand{\bp}{\tilde \beta^\prime}
\newcommand{\cB}{{\cal B}}
\newcommand{\cO}{{\cal O}}
\newcommand{\vecx}{\vec{x}}
\newcommand{\vecy}{\vec{y}}
\newcommand{\vecp}{\vec{p}}
\newcommand{\vecq}{\vec{q}}
\newcommand{\tr}{{\rm tr} }
\newcommand{\Tr}{{\rm Tr} }
\newcommand{\NP}{Nucl. Phys. }

\def\sst#1{{\scriptscriptstyle #1}}
\def\0{{\sst{(0)}}}
\def\1{{\sst{(1)}}}
\def\2{{\sst{(2)}}}
\def\3{{\sst{(3)}}}
\def\4{{\sst{(4)}}}
\def\5{{\sst{(5)}}}
\def\6{{\sst{(6)}}}
\def\7{{\sst{(7)}}}
\def\8{{\sst{(8)}}}
\def\ve{\varepsilon}
\def\vf{\varphi}
\def\F{\Phi}
\def\e{\epsilon}
\def\wg{\wedge}

\newcommand{\tamphys}{\it Center for Theoretical Physics,\\
Texas A\&M University, \\College Station, Texas 77843}

\newcommand{\auth}{I.Y. Park, A. Sadrzadeh and T.A. Tran}

\begin{document}
\begin{flushright}
\hfill{CTP TAMU-29/00}\\
\hfill{hep-th/0010116}\\
\hfill{}\\
\end{flushright}

\vspace{20pt}

\begin{center}
{\large {\bf Super Yang-Mills Operators from the D3-brane Action in a
Curved Background}}

\vspace{30pt}
\auth

{\tamphys}
     
\vspace{40pt}
\begin{abstract}
A consistent truncation of IIB on $S^5$ has been
obtained in the sector of the metric and the 4-form potential.   
The ansatz contains 20 scalars and all 15 gauge fields of 
${\cal N} = 8$ gauged supergravity in five dimensions. 
With this fully non-linear ansatz, the calculations for $n$-point correlators
of super Yang-Mills (SYM) theory via AdS/CFT are simpler than 
those in the literature that use
the {\em linear} ansatz followed by non-linear field
redefinitions. We work out the SYM operators that couple to the
scalars by expanding the Dirac-Born-Infeld (DBI) action plus
Wess-Zumino (WZ) terms around  an AdS$_5 \times S^5$ background
with the metric fluctuations. The resulting operators agree
with those based on a superconformal symmetry argument. We discuss the
significance of our results.
\end{abstract}

\end{center}

\newpage

\section{Introduction}

   Recently, a consistent fully non-linear reduction of IIB
supergravity on $S^5$ has been obtained~\cite{clpst,clp} for the
$SL(2,R)$-singlet sector comprising the metric and the 4-form
potential. In particular, the ansatz contains the scalars $T_{ab}$ in
the 20' representation of $SO(6)$ and the 15 gauge fields $A^{ab}_\1$,
of ${\cal N} = 8$ five-dimensional gauged supergravity.

The advantage of having a non-linear ansatz becomes obvious when it
comes to computing SYM correlators via AdS/CFT: it is no longer
necessary to introduce the non-linear field redefinitions that
appeared in the literature as a consequence of using a linear ansatz.
Another application of the ansatz we discuss here is to the
determination of SYM operators that correspond to given supergravity
modes.

In AdS/CFT \cite{mal,gkp,w1}, there are two known ways to determine
the SYM operators that correspond to given supergravity modes. In the
first approach \cite{w1,flz}, one considers the representations 
of the super Lie
algebra, $SU(2,2|4)$, of the fields in SYM theory and IIB 
supergravity  respectively, and matches the supergravity modes with the
SYM operators by comparing the various quantum
numbers. The other approach, which we will follow in this article, was  
proposed by Das and Trivedi \cite{dt}, where they considered the lowest
KK mode of the NS-NS 2-form fields polarized along the D3-brane world
volume. They worked out the corresponding SYM operators by
expanding the DBI action plus WZ terms around an AdS$_5 \times S^5$ 
background. They noted that the expansion arund this background, 
as opposed to a flat background, is crucial in order to obtain
the correct SYM operators. A similar method was used in \cite{dm,cgkt}

The relevance of curved backgrounds in AdS/CFT was already noticed in
\cite{ffz,lt} and was further motivated in \cite{iyp}. It is one of
the aims of the present work to consider another supergravity mode, and
to work out the SYM operators by following steps that are analogous to
those in \cite{dt}.

The rest of the paper is organized as follows. In section 2, we set
the gauge fields $A^{ab}_\1$ to zero for the purposes of concrete
computations, although keeping them does not cause any essential
additional complications. The conditions $A^{ab}_\1=0$ then induce
fifteen constraints on the scalars $T_{ab}$, since setting the
Yang-Mills fields to zero implies that the scalar currents that would
excite them must vanish. These constraints can be solved by
introducing a diagonal parameterization for $T_{ab}$, with five
independent diagonal modes.  We then compute the action for these
diagonal modes. We note that the calculations for the $n$-point
correlators are simpler than those in the literature that make use of
the linearized ansatz, and which then require non-linear field
redefinitions.  In section 3, the ans\"atze for the metric and the
4-form potential are substituted in the Abelian DBI action plus WZ
terms\footnote{To the leading order in the momentum expansion, the
contributions come only from the DBI action.}. We work out the CFT
operators by expanding them to linear order in the diagonal modes.  At
the end of the section we relax the condition $A^{ab}_\1=0$, and
obtain the CFT operators that correspond to the full set of 20 scalars
$T_{ab}$.  We conclude in section 4 with a discussion of the
significance of our results.

\section{Correlators}

 Type IIB supergravity can be consistently truncated in $D=10$ such
that only the metric and the self-dual 5-form field strength remain,
since these are the fields that are singlets under $SL(2,R)$.  The
non-linear Kaluza-Klein $S^5$ reduction ansatz for this sector has
been obtained in \cite{clpst}, and is given by
\bea
d\hat s_{10}^2 &=& \Delta^{1/2}\, ds_5^2 + g^{-2}\, 
\Delta^{-1/2}\, T^{-1}_{ab}\, 
  D\mu^a\, D\mu^b\,,\label{metans}\\
\hat H_\5 &=& \hat G_\5 + {{\hat *}\hat G_\5}\,,\label{hans}\\
\hat G_\5 &=& -g\, U\, \epsilon_{\5} + g^{-1}\, 
(T^{-1}_{ab}\, {*D}\, T_{bc})\wedge 
(\mu^c\, D\mu^a)\nn\\
&& -\ft12 g^{-2}\, 
T^{-1}_{ac}\, T^{-1}_{b\ell}\, {*F_\2}^{ab}\wedge
D\mu^c\wedge D\mu^\ell\,,\label{gans}\\
{{\hat *}\hat G_\5} &=& \fft1{5!}\, \vep_{a_1\cdots a_6}\, \Big[
g^{-4}\, U\, \Delta^{-2}\, D\mu^{a_1}\wedge \cdots \wedge D\mu^{a_5}\,
\mu^{a_6}\nn\\
&& -5 g^{-4}\, \Delta^{-2}\, D\mu^{a_1}\wedge \cdots \wedge D\mu^{a_4}
\wedge DT_{a_5 b}\, T_{a_6 c}\, \mu^b\, \mu^c \nn\\
&&- 10 g^{-3}\, \Delta^{-1}\, 
 F_\2^{a_1 a_2}\wedge D\mu^{a_3}\wedge D\mu^{a_4}\wedge D\mu^{a_5}\, 
T_{a_6 b}\, \mu^b \Big]\,,\label{gdualans}
\eea
where
\bea
&&U \equiv 2 T_{ab}\, T_{bc}\, \mu^a\, \mu^c -\Delta\, T_{aa}\,, \qquad 
\Delta \equiv T_{ab}\, \mu^a\, \mu^b\,,\nn\\
&&F_{(2)}^{ab} = dA_{(1)}^{ab} + g \; A_{(1)}^{ac}\wedge
                               A_{(1)}^{cb}\,,  \nn\\
&&
DT_{ab} \equiv dT_{ab} + g\, A_{(1)}^{ac}\, T_{cb} + g\, A_{(1)}^{bc}\, 
T_{ac}\,,\nn\\
&& \mu^a\, \mu^a = 1\,,\qquad 
D\mu^a \equiv d\mu^a +g\,  A_{(1)}^{ab}\, \mu^b\,,
\eea
and $\e_\5$ is the volume form on the five-dimensional spacetime,
whilst $\ve_{a_1\cdots a_6}$ is the tensor density in six dimensions.
$T_{ab}$ is a symmetric unimodular matrix.  The ansatz given above is
for a general $S^5$ reduction, but later we shall choose the spacetime
$ds_5^2$ to be a five-dimensional anti-de Sitter space, $AdS_5$.

   The resulting Lagrangian and the field equations for $T_{ab}$ and
$F^{ab}_\2$ are given in \cite{clpst}. For simplicity, we consider
configurations with $A^{ab}_\1=0$, which in turn enables one to use a
diagonal parameterization for $T_{ab}$, as we shall see below. The
Lagrangian is
\be
{\cal L} = R*\oneone - \fr14\,T^{-1}_{ab}*dT_{bc}\wedge
 T^{-1}_{ce}dT_{ea} - \fr12 g^2 (2T_{ab}T_{ba} - T_{aa}^2)*\oneone,
\la{lag1}
\ee
and the field equations are
\bea
T^{-1}_{c[a}*dT_{b]c} &=& 0,\nn\\
d(T^{-1}_{ac}*dT_{ba}) &=& 
- 2 g^2 (2\,T_{ac} T_{cb} - T_{ab}\,T_{cc})\epsilon_{(5)} + 
\fr13 g^2 (2\,T_{ce} T_{ec} - T_{cc}^2)
\epsilon_{(5)}. \la{eq1} \eea
Since the first equation in (\ref{eq1}) imposes fifteen constraints,
one can put $T_{ab}$ into the diagonal form
\be
T_{ab} = \mbox{diag}(X_1,X_2,X_3,X_4,X_5,X_6)\;\;\;\;
; \;\;\;\; \prod_{a=1}^6 X_a = 1.
\ee
We now adopt the parameterization used in~\cite{clp}
\be X_a = \exp\left(-\fr12 \vec{b}_a.\vec{\vp}\right), 
\ee
where $\vec{b}_a$ are the weight vectors of the
fundamental representation of $SL(6, \R)$ and 
$\vec{\vp} = (\vp_1, \vp_2, \vp_3, \vp_4, \vp_5)$ are five
independent scalars. The explicit expressions for $\vec{b_a}$ can be taken as 
follows\footnote{They also satisfy the following relations:
$ \vec{b}_i.\vec{b}_j = 8\d_{ij} - \fr4{3},\; \sum_{i=1}^6 \vec{b}_i = 0,
\; \mbox{and} \;\sum_{i=1}^6 (\vec{u}.\vec{b}_i)\vec{b}_i = 8\vec{u}
$
where $\vec{u}$ is an arbitrary vector.
}
\bea
\vec{b}_1 &=& \left(2, \fr2{\sqrt{3}}, \fr2{\sqrt{6}}, \fr2{\sqrt{10}},
\fr2{\sqrt{15}}\right),\hspace{0.3cm}  \vec{b}_2 = \left(-2, \fr2{\sqrt{3}}, 
                             \fr2{\sqrt{6}}, \fr2{\sqrt{10}},
\fr2{\sqrt{15}}\right), \nn\\
\vec{b}_3 &=& \left(0, -\fr4{\sqrt{3}}, \fr2{\sqrt{6}}, \fr2{\sqrt{10}},
\fr2{\sqrt{15}}\right), 
\vec{b}_4 = \left(0, 0, -\sqrt{6}, \fr2{\sqrt{10}},
\fr2{\sqrt{15}}\right), \nn\\
\vec{b}_5 &=& \left(0, 0, 0, -\fr8{\sqrt{10}},
\fr2{\sqrt{15}}\right), \hspace{0.9cm}
\vec{b}_6 = \left(0, 0, 0, 0, -\fr{10}{\sqrt{15}}\right).
\eea
With the diagonal form of $T_{ab}$, the Lagrangian and equations
of motion up to fourth order in $\vec{\vp}$ are\footnote{We
dropped the Einstein action treating the metric as an AdS$_5$ background.}
\bea 
e^{-1} {\cal L} &=&  - \fr12 (\pa\vec{\vp}).(\pa\vec{\vp}) +
12 g^2 + \fr14 g^2 \sum_{c=1}^6 (\vec{b}_c.\vec{\vp})^2 +
\fr1{24} g^2 \sum_{c=1}^6 (\vec{b}_c.\vec{\vp})^3\nn\\
& & - \fr5{192} g^2 \sum_{c=1}^6 (\vec{b}_c.\vec{\vp})^4 +
\fr1{128} g^2 \left[\sum_{c=1}^6 (\vec{b}_c.\vec{\vp})^2\right]^2\nn\\
&=&  - \fr12  \pa_\m\vec{\vp}\cdot\pa^\m\vec{\vp} + 12 g^2 + 2
g^2  \vec{\vp}\cdot\vec{\vp} + V_3+ V_4, \nn\\
\label{5sugra}
\Box\vec{\vp} &=& - 4 g^2 \vec{\vp} - \fr18 g^2 \sum_{a=1}^6
\vec{b}_a\,(\vec{b}_a.\vec{\vp})^2 + \fr5{48} g^2 \sum_{a=1}^6
\vec{b}_a\,(\vec{b}_a.\vec{\vp})^3\nn\\
& & - \fr14 g^2 \vec{\vp} \sum_{a=1}^6 (\vec{b}_a.\vec{\vp})^2,
\eea
where $V_3$ and $V_4$ are third-order and fourth-order
polynomials in $\vec{\f}$ respectively. Their explicit forms are
presented in the appendix. 

In \cite{lmrs}, two-point and three-point correlators for various
chiral primary operators were computed using a linear ansatz in
\cite{krv}. As a consequence of using the linear ansatz, non-linear
field redefinitions were required.  The advantage of having a
non-linear ansatz is obvious from (\ref{5sugra}): it renders such
field redefinitions unnecessary. One can easily read off two-point and 
three-point functions using the formulae in \cite{fmmr,mv}.

\section{CFT Operators from the Dirac-Born-Infeld Action}

D-branes are the objects on which open strings can end. They
also appear as solitonic solutions of supergravities and
string theory. 
Since these solutions carry mass and charge, one may have to
view the open strings as propagating in the {\em curved}
background produced by the branes to which they are
attached \cite{iyp}. The low-energy effective action for D-branes 
should then be considered in the same curved background.
  
   The authors of \cite{dt} considered the s-wave of the NS-NS $B$-field with
non-zero components along the world volume of the D3-brane.  They
noted that to obtain the correct SYM operators, it is crucial that one
expand the DBI action around an $AdS_5\times S^5$ background. In this
section, we follow \cite{dt} and expand the DBI action plus the
WZ-terms\footnote{It turns out, as we discuss below, that only the DBI
action is relevant to the leading order in the derivative expansion.}
around an $AdS_5 \times S^5$ vacuum, with the fluctuations parameterized by
the diagonal modes $\varphi_a$.  We then identify the CFT operators,
roughly speaking, as the coefficients of $\vf_a$.

Keeping only the terms linear in $\varphi$, the metric ansatz is given by
\bea
ds_{10}^2 &\simeq& \left(1-\fr{1}{4}\sum_i(\m^a)^2\vec{b}_a
                      \cdot\vec{\varphi}\right)ds_5^2 \nn\\
          & &    +\fr{1}{g^2}\left(1+\fr{1}{4}\sum_c\vec{b}_c\cdot
              \vec{\varphi}(\m^c)^2\right)\sum_a\left(1+\fr{1}{2}\vec{b}_a
                      \cdot\vec{\varphi} \right)(d\m^a)^2.
\label{metlinear}
\eea 
As previously mentioned, the ansatz quoted in
Eq.~(\ref{metans}) 
is for general $S^5$ reductions. However, we shall choose the
five-dimensional spacetime to be AdS$_5$ for our discussion. 
The structure of the D3-brane solution of type IIB supergravity in
the near-horizon region is such that the ``radius'' of the AdS$_5$ is
the same as that of the internal sphere. Therefore we
choose $ds_5^2$ to be an AdS$_5$ of radius $\fr{1}{g}$. Then we have 
\be
ds_5^2+\fr{1}{g^2}\sum_a (d\m^a)^2=g^2r^2 \sum_i(dx^i)^2
           +\fr{1}{g^2r^2}\left( dr^2+r^2\sum_a(d\m^a)^2\right).
\ee 
The metric ansatz (\ref{metlinear}) in its written form, i.e. in the
$\m$-coordinate system, does not make manifest the $SO(6)$ covariance
of the conformal field theory. A more suitable coordinate system is 
one that reveals the brane structure more
transparently; not surprisingly, such a coordinate system is of
Cartesian type,
\be
\m^a=\fr{\F^a}{r} \hspace{.5in}{\rm and}\hspace{.5in} r^2=\sum_{a=1}^6 
(\F^a)^2.
\label{nc}
\ee
Using the $\F$-coordinate system, one can rewrite
(\ref{metlinear}) in the form
\be
ds_{10}^2=g^2r^2f \, \sum_i(dx^i)^2 + \fr{1}{g^2r^2}\,\sum_{a,b=1}^6\,
g_{ab}\, d\F^ad\F^b,
\label{metric}
\ee
where 
\bea
f       &\equiv& 1-\fr{1}{4r^2}\sum_a\vec{b}_a\cdot\vec{\varphi}
                                   \,(\F^a)^2, \nn\\
g_{ab}  &\equiv&\fr{1}{g^2r^2}\left[ 
                \d_{ab}+\fr{1}{2}\vec{b}_a\cdot\vec{\varphi}\,\d_{ab}
                -\fr{1}{r^2}\vec{b}_a\cdot\vec{\varphi}\,\F^a\F^b
                +\fr{1}{4r^2}\sum_c\vec{b}_c\cdot\vec{\varphi}\,
                                   (\F^c)^2\,\d_{ab}
                 \right].
\label{fg}
\eea
The action for D3 branes in a general type IIB
background was studied in \cite{cvnw,bt,apps,mt}. 
For our discussion, it is enough to keep the metric and the 4-form
field  since the scalars come only from
these.   The relevant part of the action is
\bea
I = -\int d^4\xi \sqrt{-det(G_{ij}+{F}_{ij})}
    +\int \;\hat{C}_{(4)},
         \label{dbi}
\eea 
where  
\bea
G_{ij}        &=&\frac{\partial Z^M}{\partial\xi^i}\frac{\partial
                  Z^N}{\partial\xi^j}E^{A}_{M}E^B_N\eta_{AB}, \nn \\
C_{ijkl}      &=& C_{MNPQ} \, \pa_iZ^M\pa_jZ^N\pa_kZ^P\pa_lZ^Q.
\label{defs}
\eea
It is
the 4-form potential $C_{(4)}$ that appears in (\ref{dbi}), whereas the
ansatz in Eq.~(\ref{gans}) and (\ref{gdualans}) was obtained in terms 
of the 5-form field strength, $\hat{H}$. To obtain $C_\4$, consider 
the metric ansatz in (\ref{metans}).
With $A^{ab}_\1=0$, it becomes
\be
ds^2=\Delta^{\frac{1}{2}}ds^2_5+\frac{1}{g^2}\Delta^{-\frac{1}{2}}
            \sum_{a} X_a^{-1}d\mu^2_a,
\label{metricA0}
\ee
where
$\D=\sum_ae^{-\frac{1}{2}\vec{b}_a\cdot\vec{\varphi}}\m_a^2$. After
some algebra, one can show that the correct $C_\4$ is given by
\bea
C_{(4)} &=& \fr{1}{2g} \left( X^{-1}_a *dX_a (\m^a)^2 \right)  \nn\\
        & & - \fr{1}{4!g^4}\fr{X_{a_1}\mu^{a_1}\m^{a_2}}{\D}\;
                 \ve_{a_1\cdots a_6} d\mu^{a_3} \wg \cdots \wg d\mu^{a_6}.
\label{c4}
\eea
Acting with the exterior derivative, the first term gives $\hat{G}_\5$
up to the field equation (\ref{eq1}), which is sufficient for our
purposes.  The second term gives $*\hat{G}_\5$ plus an extra term,
which can be cancelled by adding an appropriate term to
$C_{(4)}$. However, this term is independent of $\varphi$ and can
therefore be dropped in our discussion.

We can in fact also drop the first two terms of equation (\ref{c4}), since
we shall eventually consider only the leading order in the
derivative expansion.  From now on, we may therefore
concentrate solely on the contributions coming from the DBI action.

To obtain $G_{ij}$, we substitute (\ref{metric}), (\ref{fg}) 
and  the super vielbeins obtained in \cite{mt2,kr} into the 
first equation of (\ref{defs}). For simplicity, we keep only the
bosonic fields,
\bea
G_{ij}=g^2r^2f\,\eta_{ij} + g_{ab}\,\pa_i \Phi^a \pa_j \Phi^b 
  + F_{ij} 
\eea
\noindent After some algebra, one can show that the relevant part of
the action is given by 
\bea
I[\f] 
  && =-\int  
    -\frac{g^4 r^2}{2}\sum_a\,\vec{b}_a\cdot\vec{\f}\,(\Phi^a)^2+
      \fr{1}{2}\sum_{ab}\left(\fr{1}{2}\vec{b}_a\cdot\vec{\f}\,\d_{ab}
          -\fr{1}{r^2}\vec{b}_a\cdot\vec{\f}\,\F^a \F^b\right) 
             \partial_i \Phi^a\partial^i \Phi^b \nn\\
\label{Odiag}
\eea
where the ellipses refer to the terms with higher numbers of derivatives. 

The discussion of the full 20 scalars $T_{ab}$, without imposing
$A^{ab}_\1=0$, goes very similarly, since eventually one will be 
interested only in terms that are coupled to $T_{ab}$, but do not have
any factors of $A^{ab}_\1$.   In fact the computation is almost
identical, except that one needs the new parameterization
\be
T_{ab}\equiv (e^{S})_{ab},
\ee
where $S_{ab}$ is symmetric and traceless. Using this
parameterization, one gets 
\bea
I[S_{ab}]  && =-\int\; 
    {g^4 r^2}\Phi^a \F^b S_{ab}+
       \fr{1}{2}\left(-\pa_i \F^a \pa^i \F^b
          +\fr{1}{r^2}\left[\F^a \F^c
             \partial_i \Phi^b\partial^i \Phi^c
                     +a\leftrightarrow b  \right] \right)S_{ab} \nn\\
          &&       \hspace{3in}       - \;(\mbox{trace}).
\label{O}
\eea
To leading order in the derivative expansion, we have
\be
\hspace{1.5in}I[S_{ab}] = -\int g^4r^2\left( \F^a\F^b-\fr{1}{6}\d^{ab}\F^c\F_c
                      \right)S_{ab}+\hspace{.1in}\cdots .
\label{OforS}
\ee
As one can easily show, $S_{ab}\sim \fr{1}{r^2}$ in the boundary
region, i.e. $r\rightarrow \infty$. Therefore we impose the following
boundary condition,
\be
S_{ab} \equiv \fr{1}{r^2}S^o_{ab},
\ee 
which leads to the correct CFT operator.
\be
\hspace{1in}{\cal O}^{ab}\equiv \left( \F^a\F^b-\fr{1}{6}\d^{ab}\F^c\F_c
                      \right)+\hspace{.1in}\cdots
\label{final}
\ee
\section{Conclusion}

   In this paper, we considered the 20 of scalars in ${\cal N} = 8$
$SO(6)$-gauged supergravity in $D=5$. We showed that using the
non-linear Kaluza-Klein ansatz considerably simplifies the calculation
of $n$-point CFT correlators in the AdS/CFT correspondence. Then we
substituted the ansatz into the Abelian DBI action (plus WZ terms). By
expanding the action, we could identify the CFT operators that couple
to the supergravity modes. The result is given in eq (\ref{OforS})
above, and it is in agreement with the CFT operators obtained based on
the conformal symmetry argument.

Our work, together with \cite{dt}, provides evidence for the viewpoint
taken in \cite{iyp}. It might be also viewed as in accordance with the
claim made in \cite{af}. These authors argued that the supergravity
modes are dual to the ``extended'' chiral primary operators (CPOs),
which contain, in addition to CPOs, their
descendents, since the ellipses in (\ref{final}) will include such
quantities.

One may also apply the method of \cite{dt} and the present paper to
the cases of the M2-brane and M5-brane. Another interesting
application will be to the deformations of SYM theory. So far in the
literature, one first deforms the ${\cal N}=4$ SYM theory by adding
some operators, and then tries to find the corresponding supergravity
solution. However, our results suggest that the procedure may be
reversed in those cases where a complete Kaluza-Klein ansatz is known;
one may substitute the KK ansatz into the DBI action plus WZ terms,
and then work the operators that deform the CFT
theory. We hope to report on these issues in the near future.
 
\bigskip\bigskip

\noindent {\bf Acknowledgement}\\
We thank S. Das, H. Liu, J. Rahmfeld, S. Trivedi and A.
Tseytlin for their correspondences. We also thank M.
Mihailescu and C. Pope for the useful discussions.

\newpage

\section*{Appendix : $V_3$ and $V_4$}

The cubic and quartic terms $V_3$ and $V_4$ in the scalar potential in
equation (\ref{5sugra}) are given by
\bea
V_3 &=& \fr1{45} g^2 (30\,\sqrt{3}\,\vp_1^2\,\vp_2 - 10\,\sqrt{3}\,\vp_2^3
+ 15\,\sqrt{6}\,\vp_1^2\,\vp_3 + 15\,\sqrt{6}\,\vp_2^2\,\vp_3 -
10\,\sqrt{6}\,\vp_3^3\nn\\
& & + 9\,\sqrt{10}\,\vp_1^2\,\vp_4 + 9\,\sqrt{10}\,\vp_2^2\,\vp_4
+ 9\,\sqrt{10}\,\vp_3^2\,\vp_4 - 9\,\sqrt{10}\,\vp_4^3 +
6\,\sqrt{15}\,\vp_1^2\,\vp_5\nn\\
& & + 6\,\sqrt{15}\,\vp_2^2\,\vp_5 + 6\,\sqrt{15}\,\vp_3^2\,\vp_5
+ 6\,\sqrt{15}\,\vp_4^2\,\vp_5 - 8\,\sqrt{15}\,\vp_5^3), \\
\la{cubic}
V_4 &=& - \fr1{36} g^2 (30\,\vp_1^4 + 60\,\vp_1^2\,\vp_2^2 + 30\,\vp_2^4 +
60\,\sqrt{2}\,\vp_1^2\,\vp_2\,\vp_3 -
  20\,\sqrt{2}\,\vp_2^3\,\vp_3\nn\\
& & + 30\,\vp_1^2\,\vp_3^2 + 30\,\vp_2^2\,\vp_3^2 + 35\,\vp_3^4 +
12\,\sqrt{30}\,\vp_1^2\,\vp_2\,\vp_4 -
4\,\sqrt{30}\,\vp_2^3\,\vp_4\nn\\
& & +
12\,\sqrt{15}\,\vp_1^2\,\vp_3\,\vp_4 + 12\,\sqrt{15}\,\vp_2^2\,\vp_3\,\vp_4 -
8\,\sqrt{15}\,\vp_3^3\,\vp_4 + 18\,\vp_1^2\,\vp_4^2 +
18\,\vp_2^2\,\vp_4^2\nn\\
& & +
18\,\vp_3^2\,\vp_4^2 + 39\,\vp_4^4 + 24\,\sqrt{5}\,\vp_1^2\,\vp_2\,\vp_5 -
8\,\sqrt{5}\,\vp_2^3\,\vp_5 + 12\,\sqrt{10}\,\vp_1^2\,\vp_3\,\vp_5
\nn\\
& & +
12\,\sqrt{10}\,\vp_2^2\,\vp_3\,\vp_5 - 8\,\sqrt{10}\,\vp_3^3\,\vp_5 +
12\,\sqrt{6}\,\vp_1^2\,\vp_4\,\vp_5 + 12\,\sqrt{6}\,\vp_2^2\,\vp_4\,\vp_5 +
12\,\sqrt{6}\,\vp_3^2\,\vp_4\,\vp_5\nn\\
& & - 12\,\sqrt{6}\,\vp_4^3\,\vp_5 +
12\,\vp_1^2\,\vp_5^2 + 12\,\vp_2^2\,\vp_5^2 + 12\,\vp_3^2\,\vp_5^2 +
12\,\vp_4^2\,\vp_5^2 + 42\,\vp_5^4),\nn\\
    & & +\fr{1}{2} g^2 (\vp_1^2 + \vp_2^2 + \vp_3^2 + \vp_4^2 + \vp_5^2)^2
\eea    

\end{document}